\documentclass{PoS}
\usepackage{amsmath}
\usepackage{cite}
\usepackage{url}
\usepackage{slashed}

\newcommand{\be}{\begin{equation}}
\newcommand{\ee}{\end{equation}}
\newcommand{\ba}{\begin{eqnarray}}
\newcommand{\ea}{\end{eqnarray}}
\newcommand{\bi}{\begin{itemize}}
\newcommand{\ei}{\end{itemize}}
\newcommand{\la}{\label}

\title{A new method for suppressing excited-state contaminations on the nucleon form factors}

\ShortTitle{A new method for suppressing excited-state contaminations on the nucleon form factors}

\author{Harvey B.\ Meyer, 
Konstantin Ottnad and 
\speaker{Tobias Schulz}\\
PRISMA Cluster of Excellence, 
Institut f{\"u}r Kernphysik and 
Helmholtz Institut Mainz,\\
Johannes Gutenberg-Universit{\"a}t Mainz, 
55099 Mainz, Germany \\
E-mail:  \email{meyerh@uni-mainz.de}, 
\email{kottnad@uni-mainz.de}, 
\email{schulzt@uni-mainz.de}}

\abstract{One of the most challenging tasks in lattice calculations
  of baryon form factors is the analysis and control of
  excited-state contaminations. Taking the isovector axial form factors of
  the nucleon as an example, both a dispersive representation and a calculation in chiral effective
  field theory show that the excited-state contributions become
  dominant at fixed source-sink separation when the axial current is spatially distant from the
  nucleon source location. We address this effect with a new method in which the axial current 
  is localized by a Gaussian wave-packet and apply it on a CLS ensemble with $N_f=2+1$ flavors of 
  O($a$) improved Wilson fermions with a pion mass of $m_\pi=200\,$MeV.}

\FullConference{The 36th Annual International Symposium on Lattice Field Theory - LATTICE2018\\
		22-28 July, 2018\\
		Michigan State University, East Lansing, Michigan, USA.}

\begin{document}

\section{Introduction}\noindent

The study of the electroweak form factors of the nucleon reveals
insights into its internal structure.  Electron scattering and atomic
spectroscopy can be used to probe the electromagnetic form factors
$G_E(q^2), G_M(q^2)$, whereas for the study of the axial form factors
$G_A(q^2), G_P(q^2)$, which are far less well known than the
electromagnetic form factors \cite{2002JPhG...28R...1B}, one has to
rely on weak probes, i.e.\ neutrino scattering or muon capture
processes.  Although there is a formal description of the strong
interaction given by the theory of Quantum chromodynamics (QCD), it
remains highly challenging to understand the details of low-energy
phenomena of hadronic bound states such as the proton.  To this day,
many aspects about the distribution of the proton charge and spin as
well as the contribution of the gluons to the proton spin are not
understood at a satisfactory level.

Phenomenologically, the nucleon axial charge $g_A=G_A(0)$ is known to
two parts per mille from neutron beta decay.  In comparison, lattice
QCD studies of this low-energy observable still suffer from large
statistical and systematic uncertainties~\cite{JGreenLAT18}.
% \cite{2014arXiv1412.4637G,2017arXiv170506186C}.  
One of the main
sources for the latter is the contamination of the relevant
correlation functions from excited-state contributions.  Here we
investigate a new method to overcome this problem by localizing the
axial-vector current in position space.

\section{Formalism of nucleon correlation functions}\noindent

We start with the Euclidean nucleon two- and three-point correlation functions, given by
\begin{align}
C_2(\vec{p},t_s) &= a^3 \sum_{\vec{x}}e^{-i\vec{p}\cdot\vec{x}}\,\Gamma_{\beta\alpha} 
\langle \mathcal{N}_\alpha(\vec x, t_s)\mathcal{\bar N}_\beta(0) \rangle,
% \end{align}
% \begin{align}
\\ C_{3,\mathcal{O}}(\vec{p},\vec{p}',t,t_s)
&= a^6 \sum_{\vec{x},\vec{y}}
e^{i(\vec{p}'-\vec{p})\cdot\vec{y}}
e^{-i\vec{p}'\cdot\vec{x}}
\,
\Gamma_{\beta\alpha}
\langle
\mathcal{N}_\alpha(\vec{x},t_s)
\mathcal{O}(\vec{y},t)
\mathcal{\bar N}_\beta(0)
\rangle.
\end{align}
The source-sink separation $t_s$ measures the Euclidean time between the nucleon creation (source) and annihilation (sink) point. 
A local operator is inserted at space-time position $y=(\vec{y},t)$, $0\leq t\leq t_s$, which has a nucleon matrix element given by
$\langle N,\vec{p}',s'| \mathcal{O}(0) |N,\vec{p},s\rangle = \bar u^{s'}(\vec{p}')\mathcal{O}(P,Q)u^s(\vec{p})$, $P\equiv p'+p$, $Q\equiv p'-p$.
Taking the limit $t,(t_s-t)\gg m_\pi^{-1}$ and performing a spectral decomposition, one obtains 
(NB: we use Euclidean Dirac matrices, $\{\gamma_\mu,\gamma_\nu\}=2\delta_{\mu\nu}$)
\begin{align}
C_2(\vec{p},t)
&=
e^{-E_{\vec{p}}t}|Z(\vec{p})|^2\left(1+\frac{m_N}{E_{\vec{p}}}\right) + \dots,
\\ \label{eq:C3gs}
C_{3,\mathcal{O}}(\vec{p},\vec{p}',t,t_s)
&=
\frac{Z(\vec{p}')Z^*(\vec{p})}{4E_{\vec{p}}E_{\vec{p}'}}
e^{-E_{\vec{p}'}(t_s-t)}e^{-E_{\vec{p}}t}
\,
\text{Tr}
\left(
\Gamma
(-i\slashed{p}'+m)
\mathcal{O}(P,Q)
(-i\slashed{p}+m)
\right)+\dots,
\end{align}
where the dots stand for excited-state contributions and $p_0=iE_{\vec p}=i\sqrt{\vec{p}^2+m_N^2}$, $p_0'=iE_{\vec p'}$.

It is interesting to compare the expression 
\begin{equation}\la{eq:C3hat}
\hat C_{3,\mathcal{O}}(\vec{p},\vec{p}',t,t_s) \equiv Z(\vec p')Z^*(\vec p)
\int_{-\infty}^\infty \frac{dp_0}{2\pi} \int_{-\infty}^\infty \frac{dp_0'}{2\pi} 
{\rm Tr}\Big\{\Gamma\frac{-i\slashed{p}'+m}{p'{}^2+m^2}  {\cal O}(P,Q) \frac{-i\slashed{p}+m}{p^2+m^2}\Big\} e^{ip_0'(t_s-t)+ip_0t}
\end{equation}
to Eq.\ (\ref{eq:C3gs}). Performing the $p_0,p_0'$ integrals by contour integration, the function $\hat C_{3,\mathcal{O}}$ is found 
to contain the ground-state contribution of $C_{3,\mathcal{O}}$, but it also contains additional terms due to the singularities
of the form factor in the complex plane. What makes $\hat C_{3,\mathcal{O}}$ appealing is that, for Lorentz-covariant interpolating 
operators ($Z(\vec p)$ independent of $\vec p$), it has a manifestly Lorentz-covariant form, 
which the ground-state contribution in (\ref{eq:C3gs}) does not\footnote{This observation is most easily made by  Fourier-transforming
$\hat C_{3,\mathcal{O}}$ with respect to $\vec p$ and $\vec p'$ to obtain the coordinate-space three-point function.}.
Thus it appears that $\hat C_{3,\mathcal{O}}$ is a Lorentz-covariant completion of the ground-state contribution to $C_{3,\mathcal{O}}$.

% long-distance approximation to $C_{3,\mathcal{O}}$ containing its ground-state
% contribution which respects its Lorentz-covariance properties.

Consider then the case of ${\cal O}=A_\mu^a=\bar\psi\gamma_\mu\gamma_5\frac{\tau^a}{2}\psi$ the isovector axial current, for which 
${\cal O}(P,Q)=\left[ \gamma_\mu \gamma_5 G_A(-Q^2)-i\gamma_5\frac{Q_\mu}{2m_N}G_P(-Q^2) \right]\frac{\tau^a}{2}$ 
is parametrized by the axial and induced-pseudoscalar form factors, and does not depend on $P$. To exhibit the singularities
of the form factors, we use their dispersive representation, 
\be
G_{A,P}(-Q^2) = \int_{s_0}^\infty \frac{ds}{\pi} \frac{{\rm Im\,}G_{A,P}(s)}{s+Q^2}.
\ee
Inserting these expressions into Eq.\ (\ref{eq:C3hat}), the additional terms generated can be analyzed more explicitly.
In the present case, the induced pseudoscalar form factor contains a pole at $s_0=m_\pi^2$, while the axial current has a three-pion threshold 
at $s_0=9m_\pi^2$. Keeping only the pion-pole contribution to the three-point function, 
\be
{\rm Im\,}G_P(s) = 4\pi M_N^2 g_A^\pi \delta(s-m_\pi^2) + \dots,
\ee
we obtain for the case of ${\cal O}=A_0^3$ in the proton with $\vec p'=0$ and 
$\Gamma   = \frac{1}{2}(1+\gamma_0) (1+i\gamma_5 \,\vec s\cdot\vec\gamma)$,
\ba\la{eq:C3A0pole}
&&  \hat C_{3,A_0}(-\vec{q},\vec{0},t,t_s)_{\pi~{\rm pole}} = 
%\int d^3\vec x\int d^3\vec y\; e^{i\vec q\cdot \vec y}\; {\rm Tr}\{\Gamma {\cal G}^5_0(y,x)\}_{\pi{\rm-pole}} = 
  \frac{Z(\vec 0)Z^*(-\vec q)\, M_N g_A^{\pi} \,\vec s\cdot \vec q}{E_{\vec q}}
\\ && 
\Big[ 
\frac{2E_{\vec q}\,e^{-M_Nt_s-\omega_{\vec q}\,t}}{(M_N+\omega_{\vec q})^2 - E_{\vec q}^2}
- \frac{e^{-E_{\vec q}\,t_s-\omega_{\vec q}(t_s-t)}}{\omega_{\vec q}+E_{\vec q}-M_N}
\nonumber
 + \frac{2(M_N-E_{\vec q}) e^{-E_{\vec q}\,t-M_N(t_s-t)}}{\omega_{\vec q}^2-(E_{\vec q}-M_N)^2}
\Big].
\nonumber
\ea
From the time-dependence of the last term, one identifies it with the ground-state contribution;
its denominator is nothing but $Q^2+m_\pi^2$, thus corresponding to the pion-pole contribution to $G_{\rm P}$.
The other terms are subdominant for large $t,t_s-t$; they represent $N\pi$ excited-state contributions.
% Could these terms be helpful in describing the pre-asymptotic behavior of the three-point function?

In order to confirm this interpretation, one may calculate the pion-exchange
contribution to $C_{3,A_\mu}(\vec{p},\vec{p}',t,t_s)$ in chiral
effective theory (ChEFT), using the representation $A_\mu^a= -2i f \partial_\mu\pi^a+\dots $, 
with $f$ the pion decay constant, and the interaction Lagrangian
\begin{align}
\mathcal{L}_{\text{int}}
% =\mathcal{L}_{\text{int}}^{(1)}
&=
\frac{ig_A}{f}\bar{\Psi}(x)\gamma_\mu\gamma_5\frac{\tau^a}{2}\Psi(x)\partial_\mu\pi^a(x).
\end{align}
A straightforward field-theoretic calculation then gives
\ba\la{eq:C31pix}
&&  C_{3,A_0}(-\vec{q},\vec{0},t,t_s)_{\rm ChEFT,\pi~{\rm exchange}} =  \frac{|Z|^2 g_A \,\vec s\cdot \vec q}{2E_{\vec q}}
\\ && \Big[ \frac{4E_{\vec q}(M_N+\omega_{\vec q})e^{-M_Nt_s-\omega_{\vec q}\,t}}{(M_N+\omega_{\vec q})^2-E_{\vec q}^2}
- \frac{(\omega_{\vec q}+E_{\vec q}+M_N)\,e^{-\,E_{\vec q}\,t_s-\omega_{\vec q}\,(t_s-t)}}{ \omega_{\vec q}+E_{\vec q} -M_N}
 + \frac{4M_N(M_N-E_{\vec q})\,e^{-\,E_{\vec q}\,t-M_N(t_s-t)}}{\omega_{\vec q}^2-(E_{\vec q}-M_N)^2} \Big].
\nonumber
\ea
The last term in the square bracket corresponds to the ground-state
contribution and matches exactly the third term in
Eq.\ (\ref{eq:C3A0pole}) if we identify $g_A^\pi=g_A$. The first and
second term of Eqs. (\ref{eq:C3A0pole}) and (\ref{eq:C31pix}) also
agree in the limit of small $m_\pi,|\vec q|$. In the ChEFT
calculation\footnote{See \cite{Bar:2018akl} for a complete leading-order calculation in ChEFT; 
the pion-exchange contribution is found to be dominant.}, 
it is clear that these terms correspond to excited-state
contributions.  We conclude that $\hat C_{3,{\cal O}}$ can be helpful
in describing the pre-asymptotic form of the three-point function,
especially when a pole contribution is present. Importantly,
it does not involve additional parameters as compared to the 
expression of the ground-state contribution, as it is determined by the same form factor.
% The reason is that, in a relativistic quantum theory, the presence of a pole in the form factor of a bound
% state automatically means that the particle associated with the pole
% can be excited when the current acts on the bound state.

\subsection{Current insertion in coordinate-space}

% To further analyze the origin of the excited-state contributions appearing 
% in Eqs.\ (\ref{eq:C3A0pole}) and (\ref{eq:C31pix}), we Fourier-transform $C_{3,A_0}$ 
\noindent 
Performing a Fourier transform of equation (\ref{eq:C31pix}) to localize the axial current at point $(t,\vec y)$
in coordinate space yields the following expression,
% of the correlation function in position space
\begin{align}
% C_{3,A_0}^{1\pi}(x_0,y)
\int \frac{d^3 p}{(2\pi)^3} e^{i\vec p\cdot \vec y} C_{3,A_0}(\vec p,\vec 0,t,t_s)^{\rm ChEFT}_{\pi~{\rm exch.}}
&\approx
2i|Z|^2 g_Ae^{-m_Nt_s}\vec s\cdot\vec\nabla_{\vec y}
\left[
 G_\pi(t,\vec y)-G_\pi(t_s-t,\vec y)+\frac{m_\pi^2}{2M_N}\frac{e^{-m_\pi|\vec y|}}{4\pi|\vec y|}
% G_\pi(y_0,\vec y)\Big|^{y_0=t}_{y_0=t_s-t}+\frac{m_\pi^2}{2M_N}\frac{e^{-m_\pi|\vec y|}}{4\pi|\vec y|}
\right],
\end{align}
where $G_\pi(y)= \frac{m_\pi K_1(m|y|)}{4\pi^2 |y|}$ is the pion propagator.
We have assumed the axial current to be distant from the origin $|\vec{y}|\gtrsim m_\pi^{-1}$, as 
well as $m_\pi\ll M_N$, $m_\pi^2t_s/(2M_N)\ll 1$. Thus for  $|\vec y|\gg t,t_s-t$, there is a regime where the first two (excited-state) terms 
actually dominate over the last (ground-state) term. 

The observation above motivates the idea of restricting the location of the
axial current to the region  $|\vec y|\lesssim {\rm  max}(s_0^{-1/2})$ via a `wave packet'.
An important point is that the wave-packet method does not spoil the first-principles nature of the lattice calculation:
in a finite interval of $Q^2$, the form factor
can be parametrized in a systematically improvable way via conformal-mapping techniques. 
The idea of the wave-packet method is to directly fit this parametrization 
to the three-point function with a spatially localized (axial, vector, \dots) current.
Because we want to determine the form factor primarily at low $Q^2$, we should not use 
an $\vec x$-space wave packet with a sharp edge, which will contain high-momentum modes.
A Gaussian wave-packet is then the obvious choice.

A further motivation to use a wave-packet arises for flavor singlet
currents, for which the signal-to-noise ratio is expected to be
improved, since disconnected diagrams contribute fluctuations that do
not fall off with the spatial distance (see e.g.\ \cite{Liu:2017man,Meyer:2017hjv}).

% Indeed, the standard procedure consists in computing the nucleon form factor at discrete momentum-transfer values, and 
% to then fit these data with a specific functional form of the form factor. 

\subsection{Wave Packet Method}
\noindent In the following, we describe how the wave-packet method can be applied to calculate the axial form factor
at low momentum transfer.
We introduce the ($\vec p'=0$) nucleon 3-point correlation function with a localized axial current by
\begin{align}
C_{3,\vec{s}\cdot\vec{A}}[\psi](t_s,t)
&=
a^6L^3\sum_{\vec{x},\vec{y}}\psi(\vec{y})\,\Gamma_{\beta\alpha}
\left\langle
\mathcal{N}_\alpha(x)
\left(\vec{s}\cdot\vec{A}(t,\vec{y})\right)
\mathcal{\bar N}_\beta(0)
\right\rangle =  \sum_{\vec q} \tilde\psi(\vec q) \,C_{3,\vec s\cdot \vec A}(-\vec q,\vec 0,t,t_s).
\la{eq:C3psiA}
\end{align}
On the torus, the wave packet is given by
$\psi(\vec{y})=\frac{1}{L^3}\sum_{\vec{q}}e^{i\vec{q}\cdot\vec{y}}\,\tilde\psi(\vec{q})$.
We construct a Gaussian wave packet in the plane orthogonal to the nucleon spin
by selecting $\tilde\psi(\vec{q})= \delta_{\vec s\cdot \vec q,0}\exp(-\vec{q}_\perp^2/(2\Delta^2))$,
and maintain exact projection onto zero momentum along the spin direction, in order to be 
sensitive to $G_{\rm A}$ and not $G_{\rm P}$.
We will focus on the following ratio of 3-point and 2-point correlation functions,
\begin{align}\la{eq:wtts}
w(\Delta,t,t_s)\equiv \frac{\text{Im\,}C_{3,\vec{s}\cdot\vec{A}}[\psi](t_s,t)}{C_2(\vec{0},t_s)}   
=
\frac{1}{2}\sum_{\vec{q}\in {\cal P}_\perp}
\frac{Z^*(-\vec{q})}{Z^*(\vec{0})}\left[1+\frac{m_N}{E_{\vec{q}}}\right] G_A(-Q^2)
e^{-\frac{\vec{q}^2}{2\Delta^2} -(E_{\vec{q}}-m_N)t} +\dots,
\end{align}
where ${\cal P}_\perp = \{\vec q\,|\,\vec s\cdot \vec q=0\}$ is the plane transverse to the nucleon spin.
In (\ref{eq:wtts}), the ground-state, $t_s$-independent contribution is given explicitly and the dots stand for excited-state contributions.
In this exploratory study, we use the dipole parametrization $G_A(-Q^2)=g_A/\left(1+Q^2/M_A^2\right)^2$
of the form factor, but intend to use systematically improvable
parametrizations in the near future.  Also, we fit only the
ground-state contribution to the lattice data; we have not yet
attempted to fit the form~(\ref{eq:C3hat}) or to include explicitly
excited-state terms in the fit ansatz. 
Thus the goal is to determine the parameters $(g_A,M_A)$.
We pre-determine the overlap factors $Z(\vec p)$ 
from correlated fits to the nucleon two-point correlation functions.

%  by performing correlated fits.

\section{Lattice implementation and results}\noindent
We perform an analysis of lattice data obtained in $N_{\rm f}=2+1$ QCD 
with an $\mathcal{O}(a)$-improved Wilson fermion 
action \cite{2015JHEP...02..043B} and a tree-level improved 
L\"uscher-Weisz gauge action as well as open boundary conditions in the 
time direction \cite{2011JHEP...07..036L}. 
The nucleon correlation functions were computed on the ensemble D200,
created as part of the CLS ("Coordinated Lattice Simulations") initiative. 
The parameters of the ensemble are given in Table \ref{tab:par}. Our observable
is fully O($a$) improved.
% The simulation includes four source-sink 
% separations (see Table \ref{tab:tsep}).
\begin{table}[htp]
\centering
% \begin{center}
\begin{tabular}{|c|c|c|c|c|c|c|c|}
\hline
$\beta$ & $a/\text{fm}$ & $L/a$ & $L/\text{fm}$ & $T/a$ & $m_\pi/\text{MeV}$ & $N_{\text{conf}}$ & $N_{\vec{q}}$ \\ 
\hline
$3.55$ & $0.06440$ & $64$ & $4.1216$ & $128$ & $200$ & $1021$ & $179$ \\
\hline
\end{tabular}
\caption{Simulation parameters of the D200 ensemble. The lattice spacing is taken from~\cite{Bruno:2016plf}.}
\label{tab:par}
% \end{center}
\end{table}\\
%
% \begin{table}[htp]
% \centering
% \begin{tabular}{|c|c|c|c|}
% \hline
% $16\,a$ & $18\,a$ & $20\,a$ & $22\,a$ \\ 
% \hline
% 1.03\,fm & 1.15\,fm & 1.28\,fm & 1.41\,fm \\
% \hline
% \end{tabular}
% \caption{Physical values of the different source-sink-separations.}
% \label{tab:tsep}
% \end{table}\\
% 
% This ensemble includes a set of 
We truncate the sum over $\vec q$ in Eq.\ (\ref{eq:C3psiA}) to the 
$N_{\vec{q}}=179$ lowest lattice  momenta $\vec{q}=(2\pi/L)\vec{n}$, $\vec n\in\mathbb{Z}^3$. 
The momentum $\vec p'$ of the nucleon at the sink being set to zero, the momentum transfer is $Q^2= 2m_N(E_{\vec q}-m_N)$.
As for the calculation of the gauge expectation values, we use the truncated solver method 
with bias correction \cite{BALI20101570,Blum:2012uh}.
% \begin{align}
% \langle\mathcal{O}\rangle
% =
% \left\langle
% \frac{1}{N_{\text{LP}}}
% \sum_{k=1}^{N_{\text{LP}}}\mathcal{O}_k^{\text{LP}}
% \right\rangle
% +
% \left\langle
% \frac{1}{N_{\text{HP}}}
% \sum_{k=1}^{N_{\text{HP}}}
% \left(
% \mathcal{O}_k^{\text{HP}}
% -
% \mathcal{O}_k^{\text{LP}}
% \right)
% \right\rangle
% \end{align}
%
%

\begin{figure}[htp]
\vspace{-0.4cm}
	\centering
	\includegraphics[width=0.45\textwidth]{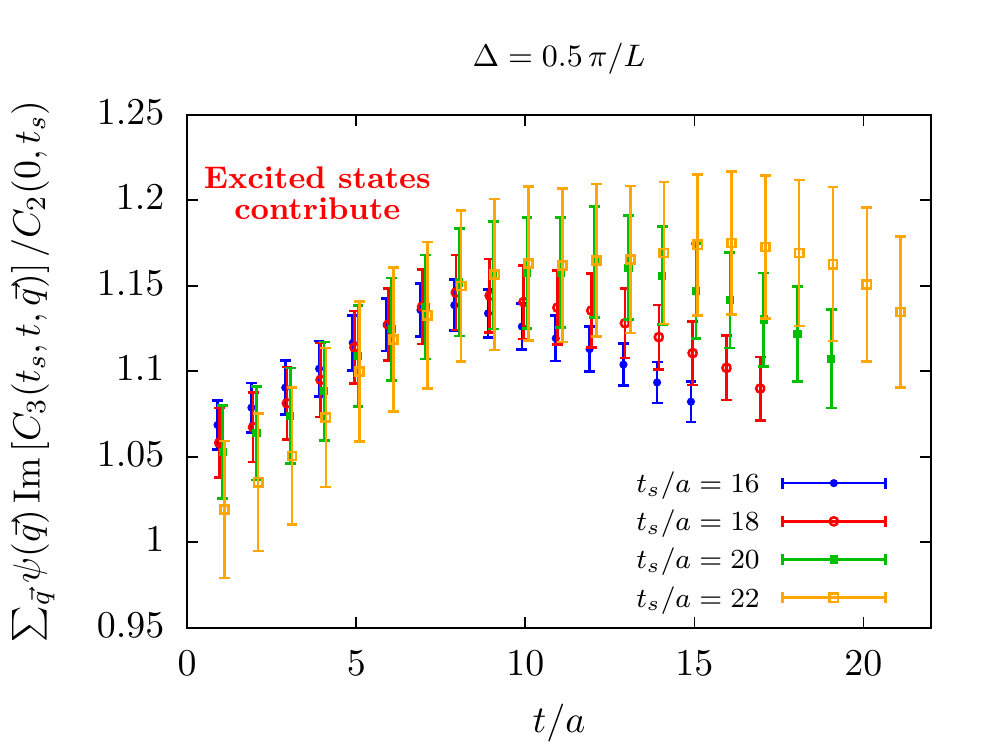}
	\includegraphics[width=0.45\textwidth]{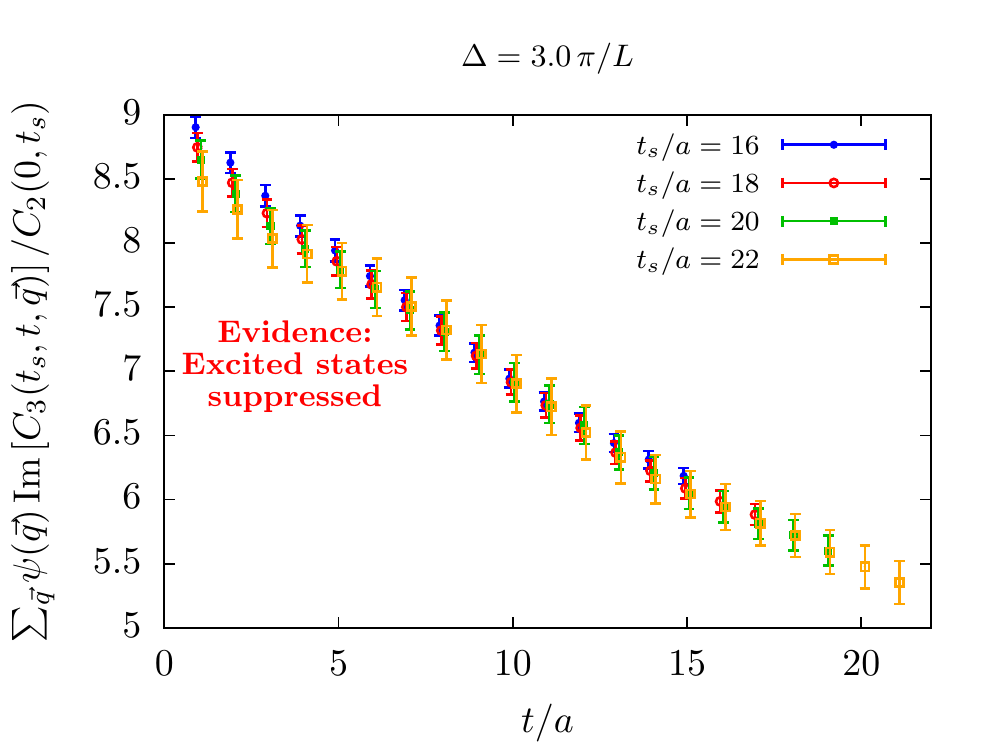}
\vspace{-0.1cm}
	\caption{The observable $w(\Delta,t,t_s)$ on ensemble D200, as a function of the
	current insertion time $t$ for two different momentum-space widths $\Delta\in\{1/2,3\}\pi/L$ of the wave packet.}
	\label{fig:wp05piL}
\end{figure}

\begin{figure}[htp]
	\centering
	\includegraphics[width=0.45\textwidth]{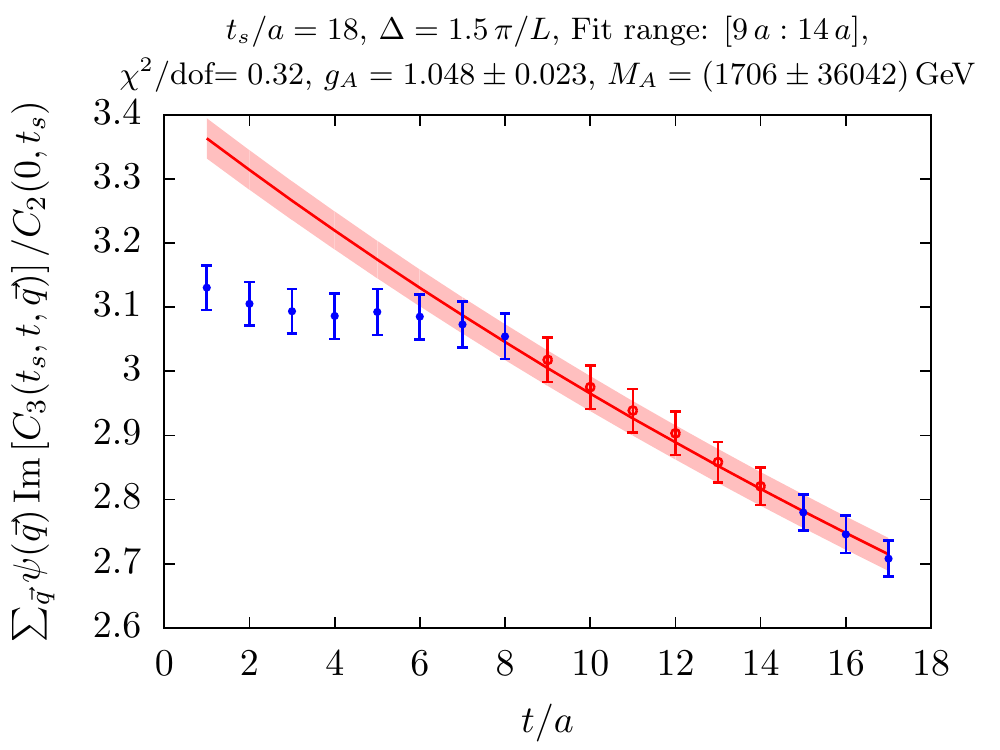}
	\includegraphics[width=0.45\textwidth]{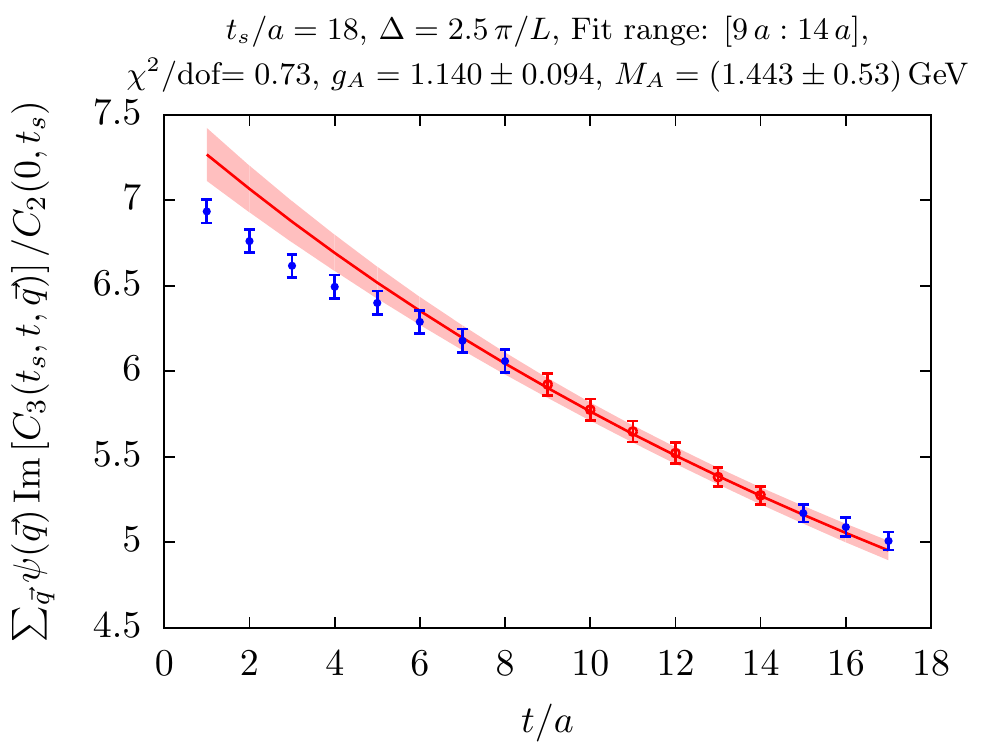}
\vspace{-0.1cm}
	\caption{Correlated two-parameter fits to the lattice data 
        for the parameters $(g_A,M_A)$, using the ground-state contribution to $w(\Delta,t,t_s)$ given in Eq.\ (2.13), % (\ref{eq:wtts}). 
         for $\Delta=1.5\pi/L$ (left) and $\Delta=2.5\pi/L$ (right).}
	\label{fig:fitSamples}
\end{figure}

\begin{figure}[htp]
	\centering
	\includegraphics[width=0.49\textwidth]{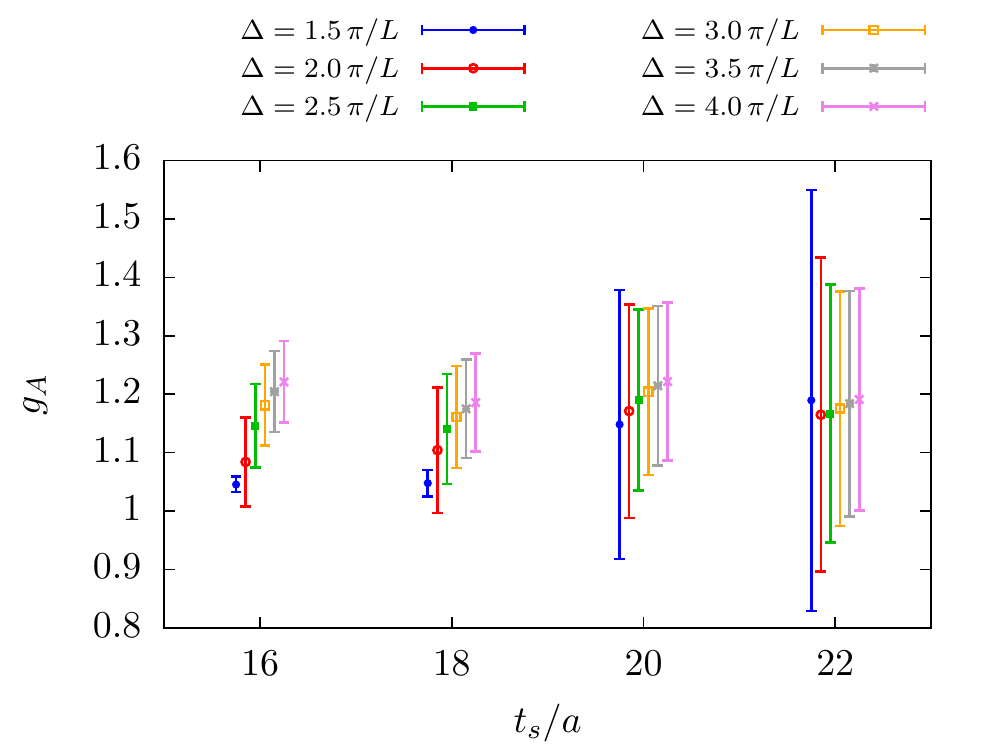}
	\includegraphics[width=0.49\textwidth]{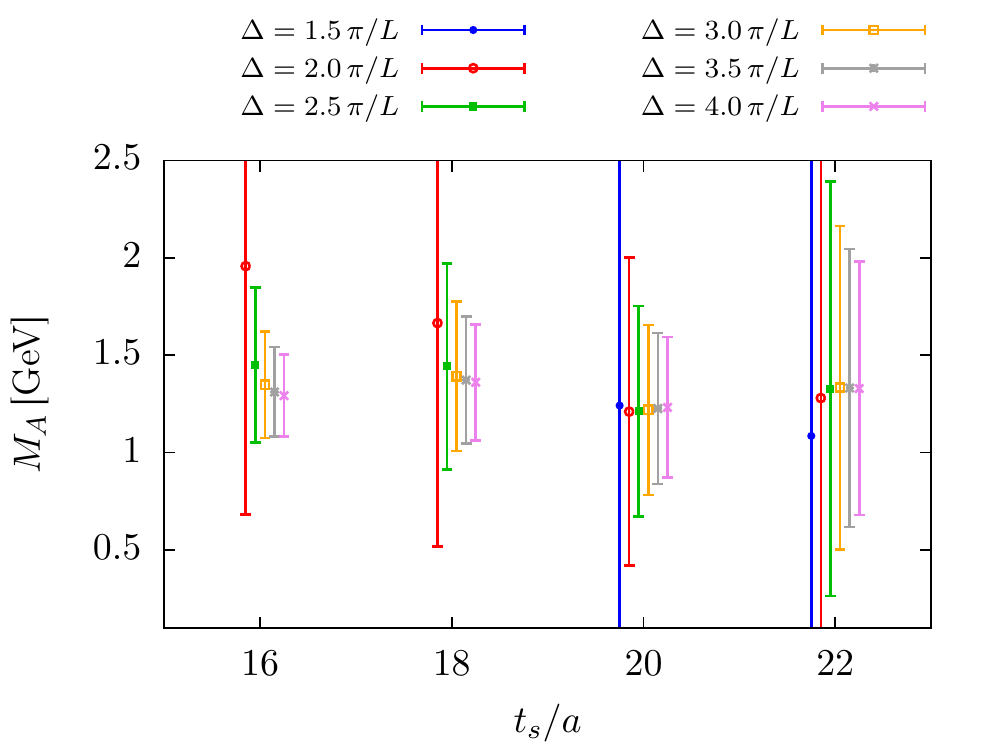}
	\caption{Summary of fit results for the axial charge (left) and the axial mass (right) for different momentum-space widths $\Delta$ of the wave packet.}
	\label{fig:resFit}
\vspace{-0.15cm}
\end{figure}

% \section{Results}\noindent
Figure \ref{fig:wp05piL} displays the first results for two different spatial localizations $\Delta=0.5\pi/L$ and 
$\Delta=3\pi/L$. 
In the left panel, a trend of $w(\Delta,t,t_s)$ as a function of $t_s$ is seen.
For the stronger localization, $w(\Delta,t,t_s)$ is independent of $t_s$ for $t\gtrsim0.4\,$fm, within the statistical uncertainty.
In Figure \ref{fig:fitSamples} we show the results of correlated two-parameter 
fits for the axial charge $g_A$ and the axial mass $M_A$ for $\Delta\in\{3/2,5/2\}\pi/L$. 
We used six time slices for the fits $t\in[9,14]\,a$. 
In the case displayed in the left panel,
the fit has no sensitivity to $M_A$, because the wave packet suppresses large-momentum contributions. 
For stronger localizations (right panel), we are sensitive to both parameters.
\\
Figure \ref{fig:resFit} shows the fit results for different sizes
$\Delta$ of localizations. The smaller source-sink separations show
smaller statistical errors but a larger sensitivity of $g_A$ to the
width of the wave packet. The results for $g_A$  are
compatible with the result ($g_A=1.188\pm 0.025$) obtained from a different method, namely 
simultaneous fits to nucleon three-point functions at $\vec q=0$ 
with different currents~\cite{2018arXiv180910638O}, but have a larger 
statistical error, as they are obtained from a small number of data.
%
% \newpage
%
%
\section{Summary and Conclusions}\noindent
We investigated a new method to compute nucleon form factors in lattice QCD
based on spatially localizing the current with a Gaussian profile.
We found evidence that for the transverse matrix elements of the axial current,
excited states get additionally suppressed as the localization becomes stronger in position space,
even though the dispersion relation of $G_A$ starts with a three-pion continuum. 
From the discussion around Eq.\ (\ref{eq:C3hat}), we expect that for form factors whose dispersive representation
starts with a two-pion cut (e.g.\ isovector vector or scalar isoscalar form factors), 
the excited-state contamination due to spatially distant contributions should be relatively 
more important, and it must be very large when the pion pole contributes.
We were able to extract the axial charge and dipole-mass from fits to lattice data
with a localized axial current in the plane transverse to the nucleon spin.
In the near future, we will perform fits to several source-sink separations simultaneously,
and apply the method to the extraction of $G_P$. Also, to better constrain 
the $Q^2$ dependence of the form factor, one could use as a set of wave packets
the Hermite polynomials times a Gaussian of fixed width.

% \subsection*{}\noindent
\noindent{\small {\bf Acknowledgements:} This project is supported by the DFG via the CRC\,1044 % ("The Low Energy Frontier of the Standard Model")
and by the European Research Council (ERC) under the European Union's Horizon 2020 research and innovation programme through grant agreement 771971-SIMDAMA.
The data has been produced on the HPC cluster "Clover" at the Helmholtz-Institute Mainz. 
The D200 ensemble was produced on JUQUEEN using computing time provided by the Gauss Centre for Supercomputing through the John von Neumann Institute for Computing.}
% We thank our colleagues for constructive suggestions, discussions and comments. 

 % \vspace{-0.1cm}
% \begin{thebibliography}{99}
% %
% %
% \bibitem{1}
% G. S. Bali, S. Collins and A. Schaefer,
% Comp. Phys. Comm. {\bf 01} (3021) 006
% [{\tt hep-lat/0910.3970}].
% %
% %
% \bibitem{2}
% V. Bernard, L. Elouadrhiri and U.-G. Mei{\ss}ner.
% \emph{J. Phys. G 28 (2002) R1}
% [{\tt hep-ph/0107088}]
% %
% %
% \bibitem{3}
% \bibitem{4}
% \bibitem{5}
% \bibitem{6}
% \bibitem{7}
% \bibitem{8}
% \end{thebibliography}

\bibliographystyle{JHEP}
\bibliography{mybibtex}

\end{document}